\documentstyle [epsfig, emulateapj]{article}
\singlespace

 1

\def\be{\begin{equation}}
\def\ee{\end{equation}}

\def\msun{{M_\odot}}

\def\etal{{\it et al.~}}

\def\HH{${\rm {H_2}}\,\,$}

\def\gs{\mathrel{\raise1.16pt\hbox{$>$}\kern-7.0pt
\lower3.06pt\hbox{{$\scriptstyle \sim$}}}}
\def\ls{\mathrel{\raise1.16pt\hbox{$<$}\kern-7.0pt
\lower3.06pt\hbox{{$\scriptstyle \sim$}}}}
\def\gtsima{$\; \buildrel > \over \sim \;$}
\def\ltsima{$\; \buildrel < \over \sim \;$}
\def\prosima{$\; \buildrel \propto \over \sim \;$}
\def\gsim{\lower.5ex\hbox{\gtsima}}
\def\lsim{\lower.5ex\hbox{\ltsima}}
\def\simgt{\lower.5ex\hbox{\gtsima}}
\def\simlt{\lower.5ex\hbox{\ltsima}}
\def\simpr{\lower.5ex\hbox{\prosima}}

\def\pp{\noindent\parshape 2 0truecm 17truecm 2truecm 15truecm}
\def\rf#1;#2;#3;#4 {\par\pp#1, #2, #3, #4. \par}

\def\pr{\ref@jnl{Phys.Rev}}     

\def\ie{{\frenchspacing\it i.e. }}

\def\href#1;#2 {{\bf #1} : {\em #2}}

\def\beq#1{\begin{equation}\label{#1}}
\def\eeq{\end{equation}}
\def\beqa#1{\begin{eqnarray}\label{#1}}
\def\eeqa{\end{eqnarray}}

\def\tento#1{\times 10^{#1}}

\def\HH{H$_2$ }
\def\H2p{H$_2^+$ }

\def\mH2p{H_2^+}



\begin{document}
\thispagestyle{empty}
\title{GRAVITATIONAL WAVE SIGNALS FROM THE COLLAPSE OF THE 
FIRST STARS}
 
\author {Raffaella Schneider\altaffilmark{1}, Andrea Ferrara\altaffilmark{2},  
Benedetta Ciardi\altaffilmark{3}, Valeria Ferrari\altaffilmark{1}, Sabino Matarrese\altaffilmark{4}
\altaffilmark{5}}

\altaffiltext{1}{Dip. di Fisica `G. Marconi', Universit\'a
degli Studi di Roma `La Sapienza' and Sezione INFN, ROMA1, P.le    
A. Moro 5, 00185 Roma, Italy}
\altaffiltext{2}{Osservatorio Astrofisico di Arcetri, L.go E. Fermi 5, 
	  50125 Firenze, Italy}
\altaffiltext{3}{Universit\`a degli Studi di Firenze, Dipartimento di
                    Astronomia, L.go E. Fermi 5, 50125 Firenze, Italy}
\altaffiltext{4}{Dipartimento di Fisica `Galileo Galilei',
Universit\'a degli Studi di Padova and Sezione INFN PADOVA, via
Marzolo 8, 35131 Padova, Italy}
\altaffiltext{5}{present address: Max-Planck-Institut f\"ur
Astrophysik, Karl-Schwarzschild-Strasse 1, D-85748 Garching, Germany}

\slugcomment{to be submitted to the ApJ }
\received{_______________}
\accepted{_______________}

\begin{abstract}
We study the gravitational wave emission from the first stars which are
assumed to be Very Massive Objects (VMOs). We take into account
various feedback
(both radiative and stellar) effects regulating the collapse of objects
in the early universe and thus derive the VMO initial mass function
and formation
rate. If the final fate of VMOs is to collapse, leaving 
very massive black hole remnants, then the gravitational waves
emitted during each collapse would be seen as a stochastic
background. The predicted spectral strain amplitude in a critical
density Cold Dark Matter universe peaks in the frequency range  
$\nu \approx 5 \times 10^{-4}-5 \times 10^{-3}$~Hz where it has a value in the range 
$\approx 10^{-20}-10^{-19}$~Hz$^{-1/2}$, 
and might be detected by {\it LISA}. The expected emission
rate is
roughly 4000 events/yr, resulting in a stationary, 
discrete sequence of bursts, \ie a
shot--noise signal.
\end{abstract}
\keywords { galaxies: formation - intergalactic medium - 
gravitational waves - cosmology: theory}

\section{INTRODUCTION}

Hierarchical models of cosmic structure formation  
predict that the first collapsed, luminous objects 
(often referred to as  Pop~III)
should form at redshift $z\approx 30$ and have a total (\ie dark +
baryonic) mass $M \approx
10^6 M_\odot$ (Couchman \& Rees 1986, Ciardi \& Ferrara 1997, Haiman 
\etal 1997, Tegmark \etal 1997, Ferrara 1998, Ciardi \etal 1999 [CFGJ], Nishi
\& Susa 1999). 
These properties are typically derived by requiring
that the cooling time, $t_c$, of the gas is shorter than the Hubble
time, $t_H$, at the formation epoch, but as we will see later 
(see CFGJ for a thorough discussion) several feedback
effects could modify this conclusion. Particularly important is the
correct treatment of the molecular hydrogen formation/destruction
network, this molecule being the only  efficient coolant  for objects
close to the above mass scale. 

As the collapse proceeds, the gas density increases and the first
stars are likely to be formed. 
However, the final product of such star formation
activity is presently quite unknown. This uncertainty largely depends on
our persisting ignorance on the fragmentation process and on its relationship 
with the thermodynamical conditions of the gas.
Ultimately, this prevents firm conclusions
on the mass spectrum of the formed stars or their IMF to be drawn. 
In the last two decades this problem has been tackled intermittently
(Silk 1977; Kashlinsky \& Rees 1983; Palla, Salpeter \& Stahler 1983; 
Carr \etal 1984, Carr 1994, Uehara \etal 1996, Omukai \& Nishi 1998). 
These studies, however, could not converge to the same conclusion 
on the typical mass range of newly formed stars in
the first protogalactic objects. Roughly speaking, two possibilities
have been proposed: either (i) Very Massive Objects
(VMOs), i.e., single stars with mass in the range $10^2-10^5 M_\odot$ (or
even larger) could be formed, or (ii) a more common
stellar cluster, slightly biased towards low-mass stars, would emerge (or some
combination of the two involving low-mass star coalescence to form a
VMO).
Early studies (Fricke, 1973, El Eid \etal 1983; Ober \etal 1983)
of the physics of VMOs were left almost unfollowed in the literature, 
probably because observational evidences of the VMO hypothesis
were lacking in the local universe. The field has recently been
rejuvenated by observations, as the Pistol Star (Figer \etal 1998)
a VMO with mass $\simeq 250 M_\odot$ and about 1 Myr old,  and
calculations (collected in the review by Larson 1998)
indicating that at earlier times the IMF was top-heavy and that 
VMOs could be a plausible outcome of the process. This possibility
would bear tremendously important consequences for the reionization
and metal enrichment of the intergalactic medium, as well as for galaxy 
formation and the nature of the dark matter.

We propose here a test of the VMO hypothesis based on the detection
of the gravitational waves (GW) emitted during the collapse into 
very massive black holes in the late phases of their evolution. 
The extreme assumption is made that the stellar population of Pop III
objects is entirely made of VMOs with mass proportional to their parent
object. This allows to estimate an upper limit to the cumulative GW 
signal from VMOs and
to compare it with existing or forthcoming experimental apparatus
(VIRGO, LIGO, LISA). 
This proposal is an ideal  development of the original suggestion    by
Thorne (1978) and Carr, Bond \& Arnett (1984) made possible by an improved
understanding of structure formation, properties of early objects 
and their GW emission mechanisms. The method adopted here presents some
similarities to the one 
outlined in Ferrari, Matarrese \& Schneider (1999), although that work
concentrated on the stochastic GW background produced by the
core-collapse of standard supernovae at relatively low redshift.

\section{THE FIRST STARS}
\label{star}

The gas in a forming galaxy is initially virialized  
in the potential well of the parent dark matter halo and 
ignition of star
formation is possible only if the gas can efficiently cool and loose
pressure support. For a plasma of primordial
composition at temperature $T< 10^{4}$ K, the typical virial temperature
of the early bound structures, molecular hydrogen is the only
efficient coolant. Thus, a minimum \HH fraction 
$f^{min}_{H_2}\approx 5 \tento{-4}$ is required for the gas to   cool  
(Tegmark \etal 1997). As this value 
is typically more than 100 times the intergalactic relic H$_2$ abundance, 
it is necessary to study in detail the \HH formation efficiency during the
collapse phase. 
The condition $f_{H_2}> f^{min}_{H_2}$ is met only by relatively large 
halos implying that for each virialization redshift there will exist
some critical, redshift dependent mass, $M_{crit}$, such that only 
protogalaxies with total mass $M>M_{crit}$ can eventually form stars.
We can then associate to each Pop III object with $M > M_{crit}$, 
a corresponding baryonic collapsed mass equal to $M_{b}=\Omega_{b} M$.
Throughout the paper we adopt a value of the baryon density parameter 
$\Omega_{b}=0.06$.

\begin{figure*}[t]
\psfig{figure=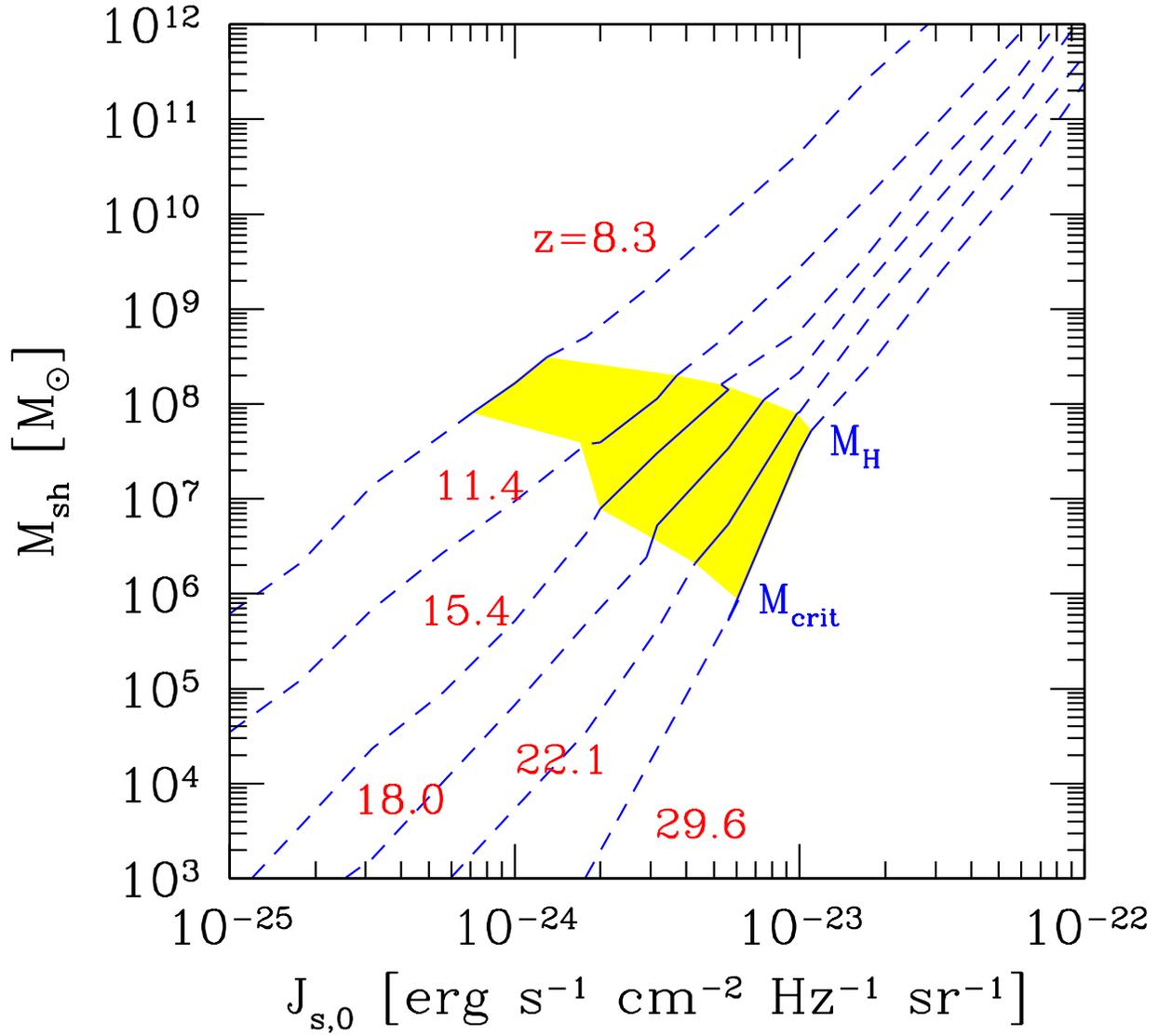}
\caption{\label{fig1}\footnotesize{Minimum total mass for
self-shielding from an external
incident flux with intensity $J_{s,0}$ at the Lyman limit. The curves
are for different redshift: from the top to the bottom $z=$8.3, 11.4,
15.4, 18.0, 22.1, 29.6. 
Radiative feedback works in the shaded area
delimited by the mass scales $M_H$ and $M_{crit}$ (see text).}}
\end{figure*}

However, this is only half of the story. In fact, 
photons from the first stars with energies
in the Lyman-Werner band and above the Lyman limit,
respectively, can photodissociate H$_2$ molecules and ionize H and He atoms
in the surrounding IGM. This is the so-called {\it radiative feedback}  
which suppresses the formation of objects more massive than $M_{crit}$
but with mass below $M_{sh}$. The latter mass scale corresponds to 
the minimum total mass required for an object to self-shield from
an external flux of  intensity $J_{s,0}$ at the Lyman limit.
The dissociating flux is the sum of two separate contributions: the first is 
coming from the background radiation produced by all luminous sources
at a given redshift; the second comes instead from the direct flux of
neighbor objects. The relative importance of the two depends on
cosmic epoch. The detailed calculation of $M_{sh}$ is rather complicated
but it is fully described in CFGJ; hence we do not
repeat it here. 
Proto-galaxies with virial temperatures $T_{vir} \simgt T_H=10^4$~K,
corresponding to a mass $M_H$=$4.4
\tento{9} M_{\odot} \, (1+z_{vir})^{-1.5} h^{-1}$,
where $z_{vir}$ is the redshift of virialization, for which cooling via
Ly$\alpha$ line radiation is possible are not affected by the radiative
feedback and are assumed to form stars unimpeded. 
These results are graphically shown in Fig. \ref{fig1}. For example, at $z
\approx 15$, it is $6\times 10^6 M_\odot \approx M_{crit} < M_{sh} < M_H 
\approx 10^8 M_\odot$, depending on the value of  $J_{s,0}$.

However, those results should be applied with a caveat to the
presently analyzed situation in which the luminosity in the early 
universe - and hence the radiative feedback - is dominated by VMOs
rather than by standard stars with a Salpeter IMF as assumed by CFGJ. 
The main problem is that the emission spectrum of VMOs
in not known (although it is currently under investigation 
[Chiosi, private communication]) and
therefore the feedback effect cannot be calculated entirely self-consistently.
To alleviate the problem, though, we note that CFGJ                     
concluded that the results are poorly sensitive to the exact form of
the spectrum as long as it is of a soft, stellar type.   

\begin{figure*}[t]
\psfig{figure=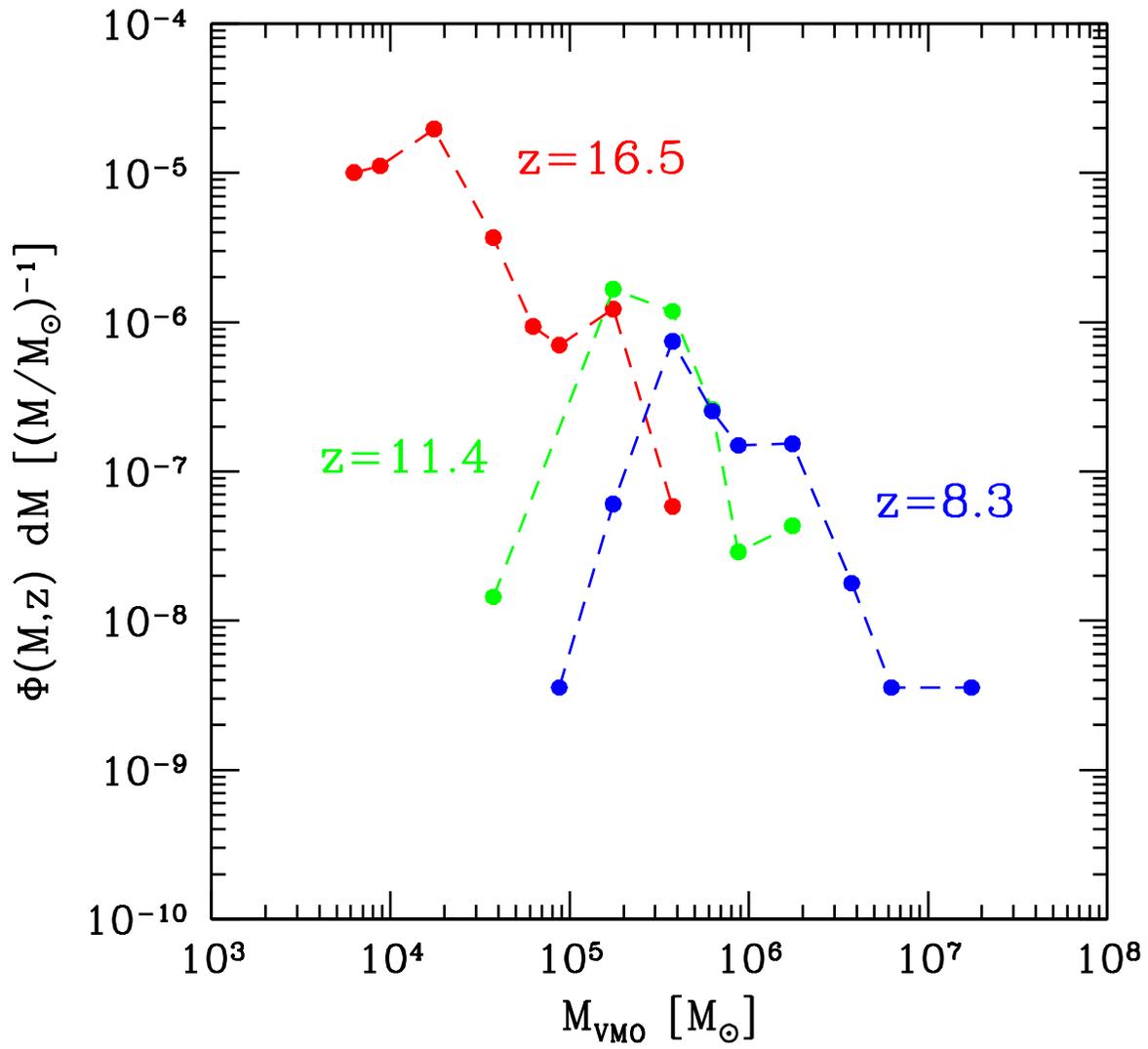}
\caption{\label{fig2}\footnotesize{VMO initial mass function evolution 
at the three different redshifts $z=16.5, 11.4, 8.3$.    
}}
\end{figure*}

In Fig. \ref{fig2} we show the evolution of the VMO
initial mass function, $\Phi(M,z)$, deduced from the simulations by 
CFGJ in which all above
effects have been included, for three different redshifts. 
These results have been obtained for a critical density Cold Dark
Matter (CDM) model ($\Omega_0$=1, $h$=0.5 with $\sigma_8$=0.6 at $z$=0);
as a consequence all the results presented here refer to the same
cosmological model.
The mass of the
VMO is here determined as $M_{VMO}=f_{b\star} M_b$, with a baryon-to-star
efficiency conversion $f_{b\star}=0.012$. As expected the
mass distribution shifts towards larger masses with time, but smaller masses
cannot form because of the collapse conditions imposed. Together with the
VMO IMF, we can directly calculate their formation rate from the curves given
in Fig. 11 of CFGJ which we do not repeat here.

A great deal of uncertainty obviously remains on the upper mass limit
of VMOs. As the mass of the parent Pop III object becomes
larger, it is increasingly difficult for the gas to collapse preventing 
fragmentation into lower mass stars (see however Silk \& Rees 1998). 
Nevertheless, conditions could be suitable 
for the steady growth of a massive star through collisions with other intermediate
mass stars. There is now substantial amount of theoretical work on this
subject, mostly for the formation of relatively massive stars ($M > 100
M_\odot$) in the local universe and neglecting the effects of a massive
dark matter halo as the one in which Pop III objects are embedded 
(Bonnell, Bate \& Zinnecker 1998, Portegies Zwart \etal 1999).
The central VMO in the Pop III cluster is then ``rejuvenated'' by each 
new collision, and its lifetime is extended considerably as a consequence.
When does this VMO mass build-up process come to an end? The usual argument 
based on the idea that radiation pressure from the VMO finally removes the  
gas producing the cluster expansion with consequent decrease of the stellar 
collision rate is probably not appropriate in this context as the gravitational
field (dominated by the dark halo) would be only very weakly affected.  
Also, radiation pressure might have been much less important in the absence
of heavy elements. 

What are the possible fates of VMOs? The answer depends essentially on their
mass. Here we are interested in objects with $10^3 M_\odot \simlt M_{VMO} 
\simlt 10^7 M_\odot$, \ie the span of the IMF in Fig. 2. Stars more massive
than about 100 $M_\odot$ are pair-production unstable (Fowler \& Hoyle 1964).
This process may lead to (Portinari \etal 1998) {\it i)}  violent pulsation 
instability with final iron core instability, {\it ii)} complete thermonuclear
explosion, or {\it iii)} collapse to a black hole. Case {\it iii)}, the one
of interest here, occurs for masses $M\simgt 200 M_\odot$ (but rotation might
increase this value, Glatzel \etal 1985). At higher masses ($M \simgt 10^5
M_\odot$) the evolution depends on metallicity, $Z$ (Fricke 1973, 
Fuller \etal 1986):
if $Z \simlt 0.005$ the star collapses to a black hole due to post-Newtonian
instabilities without ignition of the hydrogen burning; for higher 
metallicities it explodes since it could generate nuclear energy more 
rapidly from the
$\beta$-limited cycle. The former case appears to be appropriate here, 
as the metallicity
level produced by reionization is only $Z \approx 6\times 10^{-6}$
(CFGJ), and likely 
to be even smaller if the nucleosynthetic products are swallowed by black
holes. These conclusions are based on the detailed simulations by Fuller \etal
(1986) which extend up to stellar masses $M=10^6 M_\odot$. 
It has to be pointed out
that if some fraction of dark matter is present (at the level of about
0.1\%-1\%
of the VMO central density), the onset of post-Newtonian instability can be
delayed and the hydrogen burning ignited; however, this {\it favors} the
collapse to a black hole rather than the explosion, as shown by  
McLaughlin \& Fuller (1996). 
Above $10^6 M_\odot$ the study is tremendously complicated by the necessity of
taking into account general relativity effects which can influence 
the stability
and evolution of these stars. Little is know about supermassive objects 
although promising investigations are underway (Baumgarte \& Shapiro 1999,
Baumgarte, Shapiro \& Shibata 1999). In order not to add
additional uncertainty sources to our calculation {\it our main results are 
limited to VMOs with $M \le 10^6 M_\odot$}. However, because of the interesting
nature of these objects based on 
the preliminary findings of Baumgarte \& Shapiro (1999), we will also discuss 
separately the GW signal produced by the largest objects present in the derived
VMO IMF.

\section{GW EMISSION FROM VMBH COLLAPSE}

The properties of the gravitational radiation emitted
during the stellar collapse to a black hole have been extensively 
investigated during the past 20 years (see Ferrari \& Palomba
1998 for a recent review). The gravitational energy is released in a short
initial broad-band burst with efficiency $\epsilon_g$ so that the total
gravitational energy emitted is $\epsilon_g M_B c^2$, where $M_B$ is the
mass of the newly formed hole (Thorne 1986). The values of $\epsilon_g$ 
found  both in pertubative approaches and in fully numerical simulations
came out quite low. For an axisymmetric collapse, the efficiency
is less than $\sim 7 \times 10^{-4}$ (Stark \& Piran 1985). However, if the  
star is rotating sufficiently rapidly to undergo a dynamical bar mode
instability prior to forming the black hole, then the energy released in
gravitational waves can be substantially higher (Smith, Houser \& Centrella
1995). 
Therefore, the collapse of a VMO promises to be a very interesting source
for gravitational wave detection.

For the sake of simplicity, we assume that the gravitational energy
released during each collapse, $\Delta E_g = \epsilon_g M_B c^2$, 
is emitted in a broad-band burst centered at a frequency, $\nu_0=c/10 R_g$,
which corresponds to a 
wavelength of order 10 times the Schwarzschild radius, 
$R_g=2\, G\, M_B / c^2$, 
associated to the hole (Thorne 1978, Carr, Bond \& Arnett 1984). 
The spectrum of gravitational waves emitted
during the collapse can be approximated to a Lorenzian,
\be
\frac{dE}{d\nu}= \frac{\Delta E_g}{\nu_0 {\cal N}}\, 
\,\frac{\nu^2}{(\nu-\nu_0)^2+\Gamma^2}   
\label{eq:single}
\ee
where $\Gamma=(2 \pi \Delta t)^{-1}$, $\Delta t=1/\nu_0$ is the typical 
duration of the burst, $\nu_0 \, {\cal N}$ is the normalization, 
\[
\nu_0 \, {\cal N}=\int_{0}^{\nu_{max}/\nu_0}\!\!\!\!
d{\tilde \nu} \frac{{\tilde \nu}^2}{(1-{\tilde \nu})^2 +0.03}
\]
with ${\tilde \nu}=\nu/\nu_0$ and $\nu_{max}=c/R_g$ the maximum frequency
emitted by the source.
While the available theoretical waveforms are too uncertain to warrant a 
more elaborate analysis, this crude approximation well highlights the main
features and assumptions of the model.

The average gravitational flux emitted by a source at a distance $r$ can
be easily shown to be,
\be
\langle \frac{dE}{d\Sigma d\nu} \rangle = \frac{1}{4 \pi r^2} 
\int d\Omega \left [\frac{dE}{d\Omega d\nu} \right] = \frac{1}{4 \pi r^2} 
\frac{dE}{d\nu}.
\ee
For sources at cosmological distances, the above expression can be 
immediately generalized to,
\be
\langle \frac{dE}{d\Sigma d\nu} \rangle = \frac{(1+z)^2}{4 \pi d_L(z)^2} 
\frac{dE_e[\nu(1+z)]}{d\nu_e},
\label{eq:mean}
\ee  
where $\nu = \nu_e \,(1+z)^{-1}$ is the redshifted emission frequency, 
$\nu_e$, and
$d_L(z)$ is the luminosity distance to the source.

\section{PREDICTIONS AND IMPLICATIONS}

If the final fate of VMOs is to collapse, leaving  
very massive black hole (VMBH) remnants, then the overall effect
of the gravitational waves
emitted during each collapse would be seen today as a stochastic background.
The signal produced by these  events can be computed integrating 
the gravitational signal contributed by each source over the differential 
source formation rate,
\be
\frac{dE}{d\Sigma d\nu dt}[\nu] = \int\int dz \, dM \, \Phi(M,z) \, 
\frac{\dot{\rho}(z)}{(1+z)} \langle \frac{dE}{d\Sigma d\nu} \rangle 
\frac{dV}{dz},
\ee
where $\dot{\rho}(z)/(1+z)$ is 
the VMO cosmic formation rate per comoving volume obtained from the 
CFGJ simulations.

The spectral energy density allows the evaluation of 
the corresponding spectral strain amplitude,
\be
S_h[\nu] = \frac{2 G}{\pi c^3} \, \frac{1}{\nu^2} \,\frac{dE}{d\Sigma d\nu dt}
[\nu].
\ee
Clearly, the relevant parameters which determine the amplitude and the
location of the signal are the efficiency $\epsilon_g$ and the fraction
of the initial mass which participates to the collapse, $\phi_B=M_B/M$.
The value of $\phi_B$, as well as its dependence on $M$, is very uncertain.
For an axisymmetric collapse, only about 10\% of the initial stellar mass 
collapses to the final black hole (Stark \& Piran 1985). However,
Baumgarte \& Shapiro (1999) suggest that for $M \simgt 10^6 \msun$,
only a few percent of the 
initial mass is left outside of the black hole, most likely in the form 
of a disk.
Thus, it is not unreasonable to assume that $\phi_B$ may vary in the range
0.1-0.9 (see also Carr, Bond \& Arnett 1984).

The stochastic background signal predicted by our model is shown in 
Fig. \ref{fig3} for $\epsilon_g=10^{-4}$ and $\phi_B=0.1$. 
There we compare the spectral energy density 
produced by pre-galactic VMBHs to the more recent contribution from
core-collapse SNe at $z<6$ (see Ferrari, Matarrese \& Schneider 1999).
Due to their larger masses, the gravitational signal from the birth 
of VMBHs falls
at much shorter wavelengths than that produced by core-collapse SNe.

In spite of their great distances, pre-galactic VMOs produce a
background signal which adds, as a confusion noise component,
to the {\it LISA} sensitivity curve in
the range $10^{-3}-10^{-2}$ Hz (see Fig. \ref{fig4}), 
for the parameters assumed in the model. 
Thus, we expect that {\it LISA} will be able to place an upper limit
on the intensity of the Pop III background signal in this frequency range.

\subsection{ Contribution from Supermassive Objects}

\begin{figure}
\psfig{figure=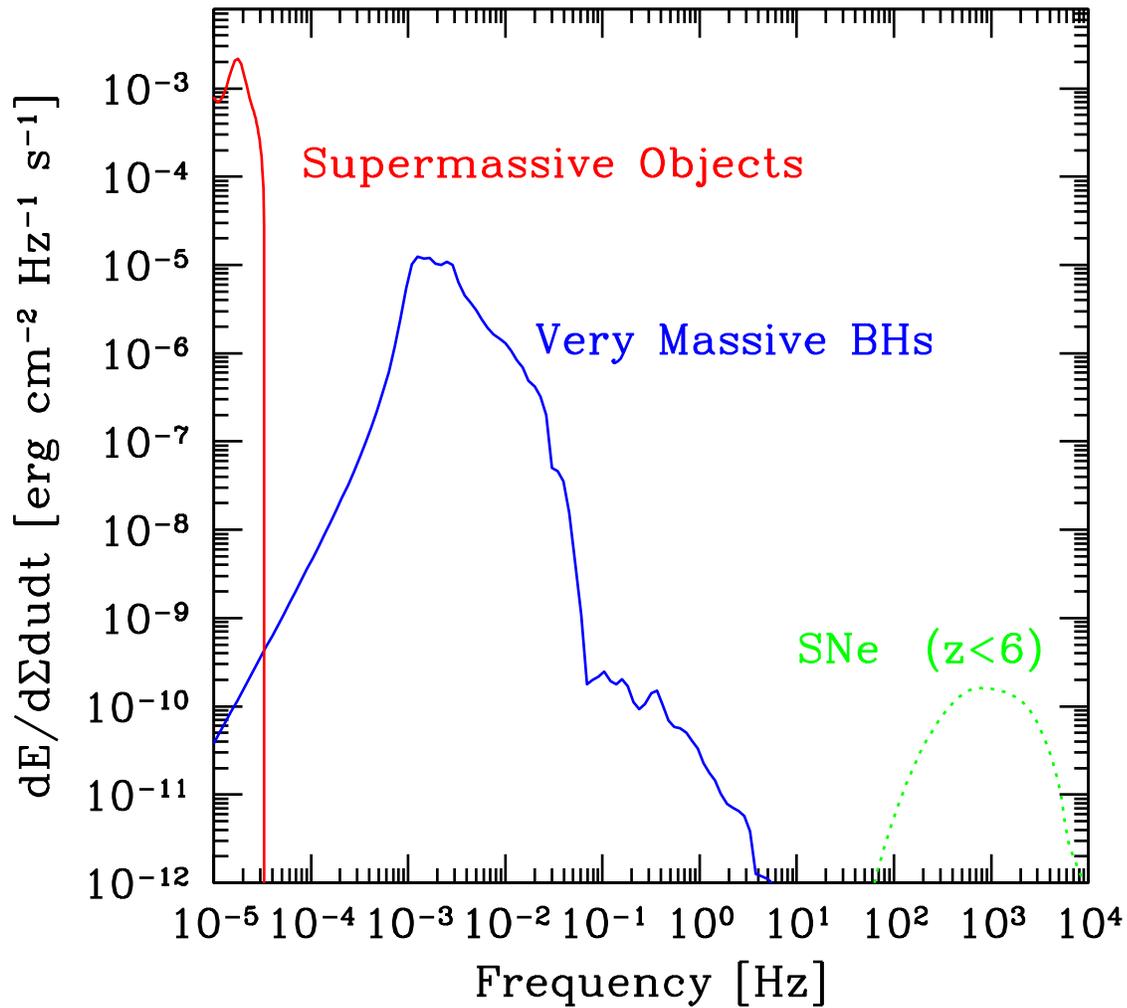}
\caption{\label{fig3}\footnotesize{
Stochastic background signal produced by the
VMBHs remnants of PopIII stars and by core-collapse SNe leaving black hole
remnants at $z<6$. The background signal produced by VMBHs is computed assuming
$\epsilon_g = 10^{-4}$ and $\phi_B=0.1$ (see text). Also shown is the
contribution from the collapse of supermassive ($M\simgt 10^6 M_\odot$)
objects.
}}
\end{figure}

\begin{figure}
\psfig{figure=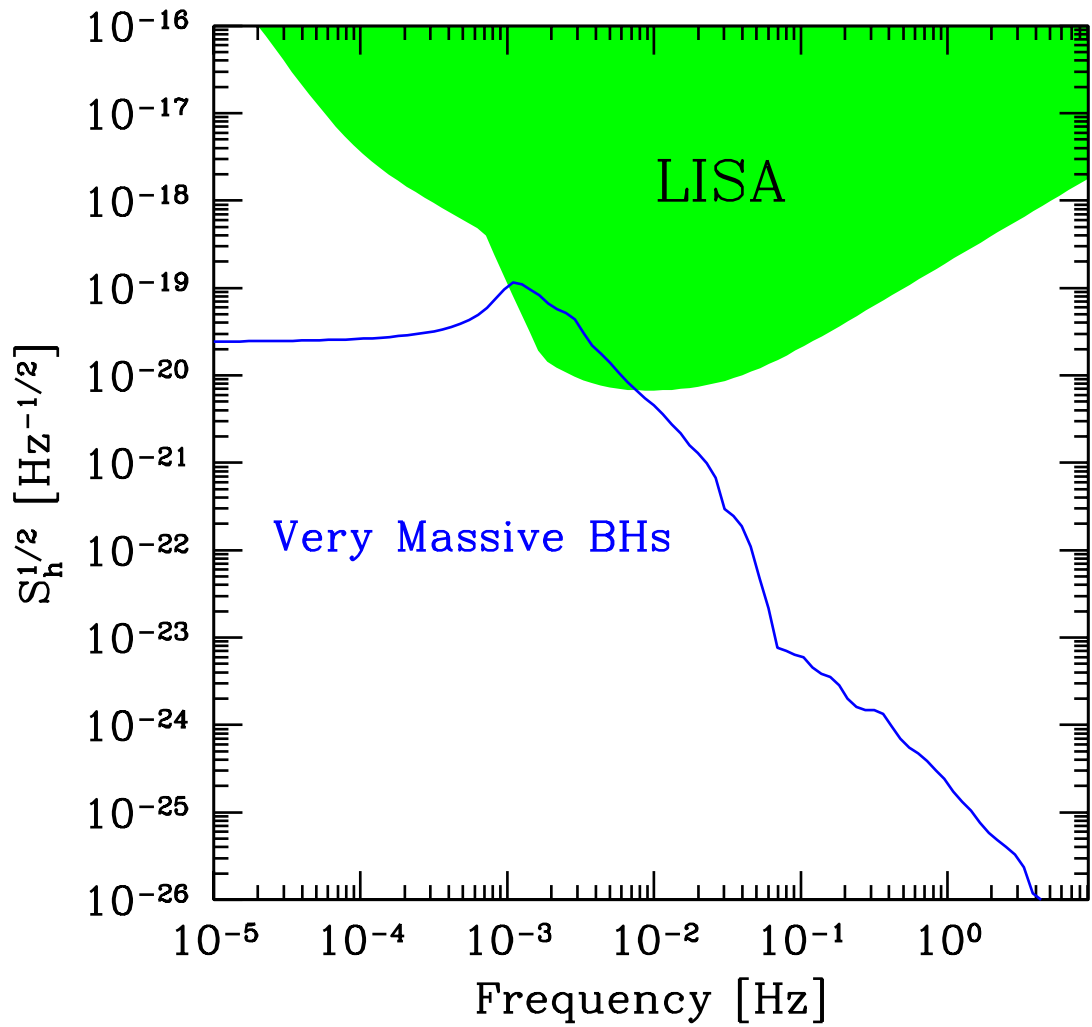}
\caption{\label{fig4}\footnotesize{
The spectral strain amplitude of the PopIII stars signal (assuming $\epsilon_g =
10^{-4}$ and $\phi_B=0.1$) is compared to the
sensitivity curve of the {\it LISA} space interferometer. The {\it
LISA} sensitivity
curve accounts for both the instrumental noise component and for the
confusion noise component due to double white dwarfs binaries (Bender \& Hils
1997).}} 
\end{figure}

In Fig. \ref{fig3} we also show a crude estimate of the signal produced
by VMOs with masses $\simgt 10^6 \msun$ which undergo a dynamical bar 
instability before the final implosion. This possibility has been recently
discussed by Baumgarte \& Shapiro (1999). They 
investigated the secular evolution of supermassive stars (SMSs) 
with masses $\simgt 10^6 \msun$, up to the onset of the instability.
The gravitational efficiency of the collapse as well as the detailed waveforms
of the resulting signal can be definitely assessed only with a numerical, 
three-dimensional hydrodynamics simulation in general relativity
(Baumgarte, Shapiro \& Shibata 1999). However, based on simple
arguments, these authors suggest that the collapsing star
may form a nonaxisymmetric bar before it forms a black hole.

Numerical simulations of the dynamical bar mode instability in compact stellar
cores with stiff equation of state ($n<1.5$) have shown that a burst of
gravitational radiation is emitted 
with an efficiency $\epsilon_g$ ranging
between $10^{-4}$ and $10^{-2}$, depending on the initial equatorial radius
of the bar, $R_{eq}$, and on the polytropic index (Houser \& Centrella 1996).
The burst is centered at a frequency $\sim 2 \nu_{bar}$, where
$\nu_{bar}$ is the rotation frequency of the bar, 
\[
\nu_{bar}=\frac{1}{2\pi} \, \left(\frac{GM}{R_{eq}^3}\right)^{1/2}.
\]  
The width of the gravitational burst depends sensitively on the 
polytropic index: stiffer models undergo several episodes of bar formation and
recontraction, emitting a sequence of bursts of decreasing amplitude.
The structure of SMSs with $M \simgt 10^6 \msun$ is that of an $n =3$ polytrope
(Baumgarte \& Shapiro 1999) and it is reasonable to
assume that the radiation emitted would be concentrated in a single burst
of width $\sim 2 \nu_{bar}$. Following Baumgarte \& Shapiro (1999) we assume, 
\[
R_{eq} \sim 1.5 R_{pol} \qquad \qquad  \qquad \qquad  
R_{pol} \sim \frac{15 G M}{c^2}
\]
where $R_{pol}$ is the polar radius at the onset of the bar instability. 
With these parameters,
we model the spectrum emitted by a VMO with $M > 10^6 \msun$ using 
eqs.
(\ref{eq:single}) and (\ref{eq:mean}) 
with an efficiency $\epsilon_g = 10^{-4}$ and $\phi_B=0.1$. 

As it can be seen from Fig. \ref{fig3}, 
these objects produce 
a significant signal at frequencies $<10^{-4}$\enspace Hz, too small to be
detectable with {\it LISA}.  
   
\subsection{ Rates and Duty Cycle}                     

\begin{figure}
\psfig{figure=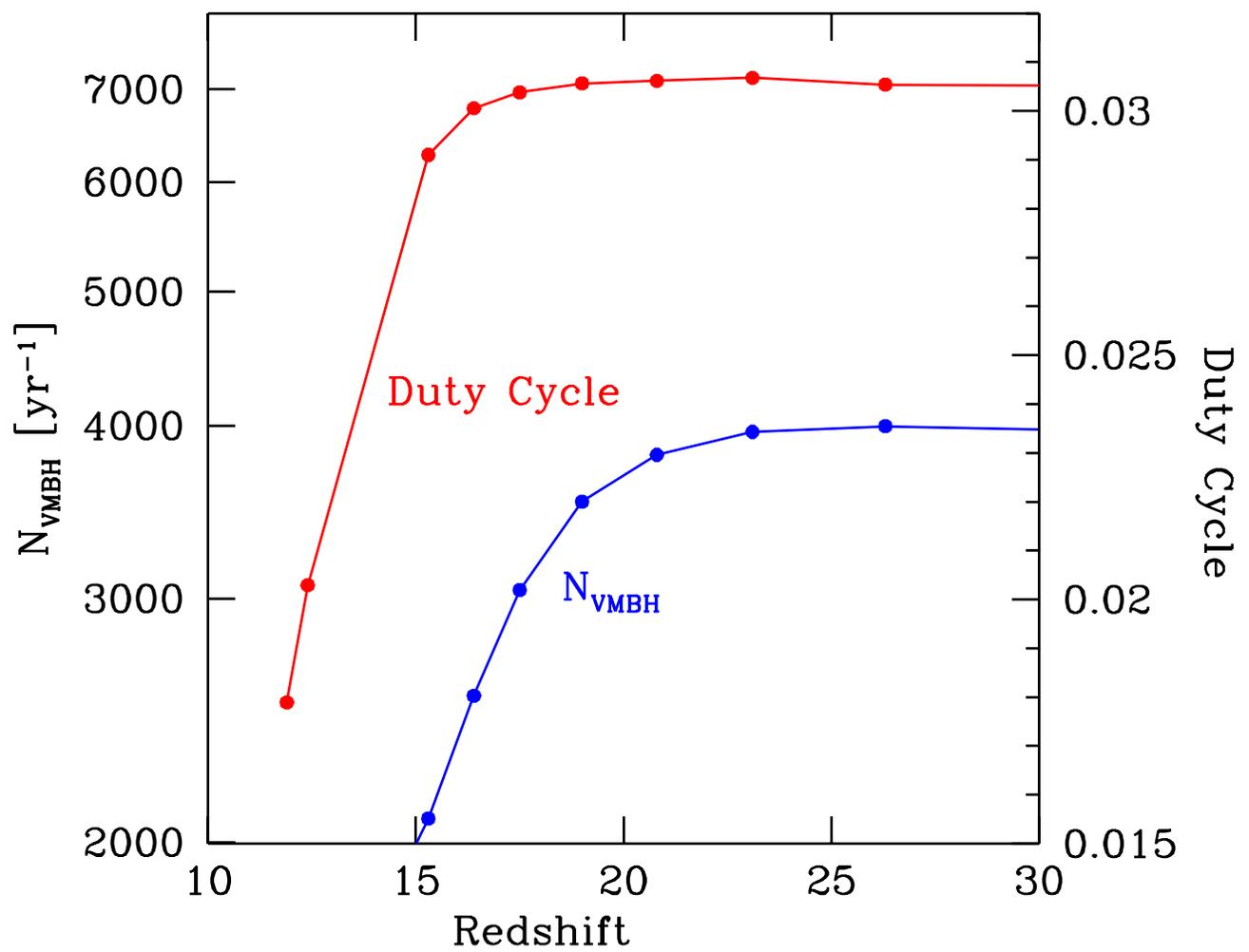}
\caption{\label{fig5}\footnotesize{ 
Rate of VMBH formation  
and duty cycle of the gravitational-wave signal produced by VMBH collapses 
as a function of redshift.}}
\end{figure}

The rate of VMBH formation is shown in Fig. \ref{fig5} as a function of 
redshift. The total number of VMBHs formed per unit time is 
$N_{VMBH} \sim 4000$ events/yr. The ratio of the duration of each
burst to the separation between successive bursts, i.e. the duty cycle, 
\be
\frac{d DC[z]}{dz}=\frac{\dot{\rho}(z)}{(1+z)} \, 
\frac{dV}{dz} \, \frac{(1+z)}{\nu_0} \, \int dM \Phi(M,z)  
\ee
is also shown in Fig.\ref{fig5} as a function of $z$. 
It is clear that the overlap
condition, $DC>1$, is not satisfied even if we consider all VMOs out to the 
farthest $z$. Thus, we find, contrary to previous claims 
(Carr, Bond \& Arnett 1984), that VMBHs that originate from Pop III stars 
do not generate an overlapping background but rather a stationary, discrete
sequence of bursts, i.e. a shot-noise signal 
(see also Ferrari, Matarrese \& Schneider 1999). 

Thus, though a stochastic background at comparable frequencies and amplitude
might have been generated in the very early universe, it would still be 
possible to disentagle any Pop III 
gravitational signature through this peculiar shot-noise character.

\section{SUMMARY AND DISCUSSION}        

We have investigated the gravitational wave emission from the collapse of
Very Massive Objects formed in the early universe. The presence of such objects
would bear tremendously important consequences for the intergalactic medium 
reionization, the generation of the first metals and for their
contribution to the dark matter in the universe.   

The predicted spectral strain amplitude in a critical density CDM
universe peaks in the  frequency range 
$\nu \approx 5 \times 10^{-4}-5 \times 10^{-3}$~Hz where it has a value in the
range $\approx 
10^{-20}-10^{-19}$~Hz$^{-1/2}$, which is above the {\it LISA} 
sensitivity curve. 
The expected emission event rate is roughly 4000 events/yr, resulting in a
stationary, discrete sequence of bursts, \ie a shot--noise signal.
The issue of the actual detectability of our signal by {\it LISA}   
is more complex, as cross-correlation techniques, which would be needed 
to disentangle a stochastic background from the noise, cannot be
applied here (see, e.g. Flanagan \& Hughes 1998). So, any background
would actually add as a confusion limited noise component to the {\it LISA}
instrumental noise. In this sense {\it LISA} will  place an
upper limit to the amplitude of our signal. 
On the other hand, 
the predicted stochastic background has a shot-noise structure,
similarly to the background produced, in the high frequency
bandwidth, by the core-collapse to black holes
in standard supernovae at $z \simlt 6$  
(Ferrari, Matarrese \& Schneider 1999, see Fig. 3).

Thus, standard detection techniques, which have been developed for
continuous stochastic signals, can be applied only if the integration
time of the antenna is much longer than the typical separation
between two successive bursts 
(of the order of a few hours for the present study).

Specific techniques should be investigated in order to assess
how far the shot-noise structure can be exploited to help the detection
or, at least, to    
distinguish the signal from the instrumental noise or from
continuous backgrounds contributing in the same band. 

Our underlying assumption that down  to $z\approx 10$ fragmentation in 
collapsing cosmological objects is inhibited by the absence of metals (and 
therefore only very massive stars are formed) is clearly strong and,
at present, untestable, but yet not an unreasonable one. Its
motivation is that 
it allows to estimate an upper limit to the expected gravitational wave 
emission from this population of astrophysical objects. The growth of 
the metallicity level in the universe not only favors fragmentation of
the gas, but also quenches the formation 
of VMBHs as these objects tend to explode, as discussed in Sec. 2. For these
reasons, the GW contribution from collapsing VMOs can only come from redshifts
higher than approximately 10.

Some details of the calculation are uncertain,
as for example the exact shape of the GW emission spectrum from a
collapsing VMO. However, a different choice of the spectrum (\ie the one
suggested by Stark \& Piran 1985) produces only a slight difference in the
integrated spectral energy distribution. Indeed, the expected signal could be
even higher than what predicted here if the conversion efficiency of the total 
gravitational energy into GWs is higher than the value $10^{-4}$ used
here, as suggested by some previous studies (the amplitude of the GW signal 
is $\propto \epsilon_g$).

Detecting a GW signal from VMOs will become possible with the new class of GW
detectors as {\it LISA}, according to our predictions. 
This would allow to directly 
test epochs and objects which would be otherwise difficult to reach with
other instruments, at best for a very long time. This experiment could also 
provide stringent limits to the cosmic star formation history at redshifts
$z>10$ and to investigate the physics and properties of the first stars in the 
universe. The remnants of such primordial objects could still be present
in halos of galaxies like our own (for an excellent review see Carr 1994). 
The type of BH remnants discussed here are interesting dark matter candidates:
gravitational lensing effects measured from the line-to-continuum 
variation of quasars suggest that the lensing objects in the
intervening galaxies
have a mass in the range $3\times 10^4 - 3\times 10^7 M_\odot$ (Subramanian \& 
Chitre 1987). However, the upper end of this interval might be constrained
to be $\simlt 10^6$ by either dynamical (heating of disk stars, Lacey \&
Ostriker 1985) or luminosity
(accretion of gas in the halos of galaxies, Ipser \& Price 1977; Hegyi \etal
1986) constraints. Interestingly enough, this interval appears not only to 
well match the predicted VMO IMF (see Fig. 2), but also to 
provide a GW signal which might be detectable by {\it LISA}.

\bigskip
We would like to thank S. Shapiro and K. Thorne for useful suggestions.
BC acknowledges support from a SAO Fellowship at CfA.

\newpage

\vfill
\eject
\end{document}